%% file: draft_KKenu.tex
\documentclass[aps,prd,twocolumn,showpacs,amsmath,amssymb]{revtex4-1}
\usepackage{amsmath}
\usepackage{graphicx}
\usepackage{subfigure}
\usepackage{epstopdf}
\usepackage{color}
\usepackage{multirow}
\usepackage{setspace}
\usepackage{overpic}
\usepackage{amssymb}
\usepackage{lineno}
\usepackage{bm}
\usepackage{rotating}
\usepackage{makecell}
\usepackage[utf8]{inputenc}
\usepackage{hyperref}
\hypersetup{
	colorlinks=true,
	linkcolor=blue,
	filecolor=blue,
	urlcolor=blue,
	citecolor=blue,
}
\let\oldequation\equation
\let\oldendequation\endequation

\renewenvironment{equation}
  {\linenomathNonumbers\oldequation}
  {\oldendequation\endlinenomath}

\begin{document}

\title{\bf \boldmath
Search for $D^{0}\to K_{S}^{0} K^{-} e^{+}\nu_{e}$,
$D^{+}\to K_{S}^{0} K_{S}^{0} e^{+}\nu_{e}$, and $D^{+}\to K^{+}K^{-} e^{+}\nu_{e}$}

\input{authorlist_2023-08-29}

\begin{abstract}
A search has been performed for the semileptonic decays $D^{0}\to K_{S}^{0} K^{-} e^{+}\nu_{e}$,
$D^{+}\to K_{S}^{0} K_{S}^{0} e^{+}\nu_{e}$ and $D^{+}\to K^{+}K^{-} e^{+}\nu_{e}$,
using $7.9~\mathrm{fb}^{-1}$ of $e^+e^-$ annihilation data collected at the center-of-mass energy $\sqrt{s}=3.773$ GeV by the BESIII detector operating at the BEPCII collider.
No significant signals are observed, and upper limits are set at the 90\% confidence level of  $2.13\times10^{-5}$, $1.54\times10^{-5}$ and $2.10\times10^{-5}$  for the branching fractions of
$D^{0}\to K_{S}^{0} K^{-} e^{+}\nu_{e}$,
$D^{+}\to K_{S}^{0} K_{S}^{0} e^{+}\nu_{e}$ and $D^{+}\to K^{+}K^{-} e^{+}\nu_{e}$, respectively.
\end{abstract}

\maketitle

\oddsidemargin  -0.2cm
\evensidemargin -0.2cm

\section{Introduction}

\begin{figure}[htbp]
\centering
\subfigure[]{\includegraphics[width=0.216\textwidth]{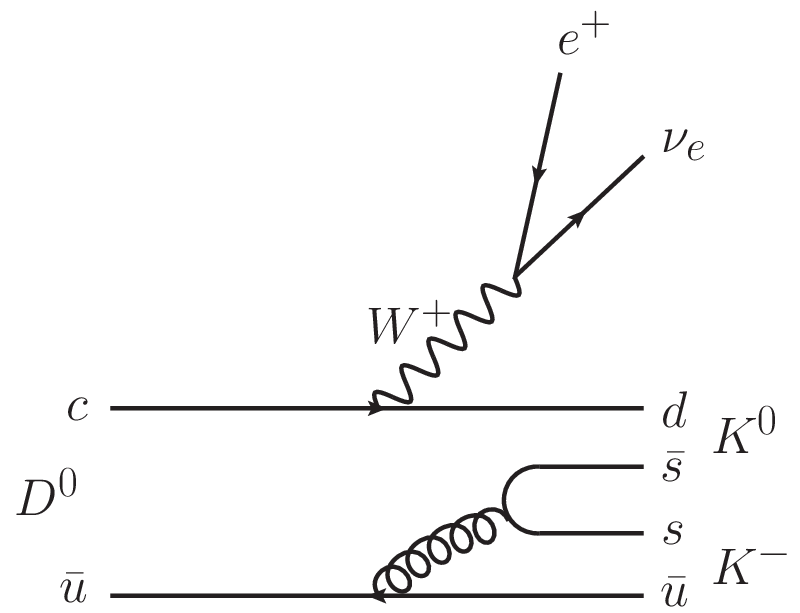}}
\subfigure[]{\includegraphics[width=0.246\textwidth]{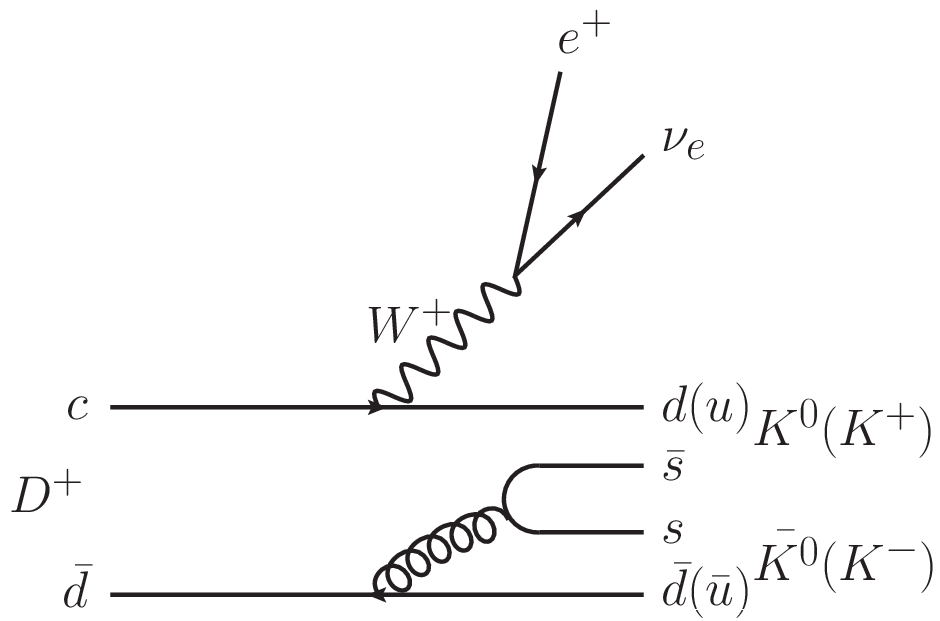}}
\caption{\footnotesize Tree-level Feynman diagrams of (a) $D^0\to a_{0}(980)^{-}e^{+}\nu_{e}$ and (b) $D^+\to a_{0}(980)e^{+}\nu_{e}$.}
\label{fig:fey}
\end{figure}

The semileptonic decays of charmed mesons offer a clean environment to explore the strong and weak interactions in the charm sector. Over the years, the semileptonic decays of $D$ mesons  into pseudoscalar and vector mesons have been investigated extensively by various experiments, such as MARKIII, BESII, CLEO-c, BaBar, Belle, LHCb, and BESIII,  and their findings are comprehensively summarized in Ref.~\cite{ref::pdg2022}. In contrast, experimental studies of semileptonic~(SL) decays involving scalar mesons are relatively limited. In 2018, the BESIII Collaboration reported the observation of the semileptonic decays $D \to a_0(980)e^+ \nu_{e}$ with $a_0(980)\to \pi\eta$~\cite{BESIII:2018sjg}, with branching fractions~(BFs) comparable to theoretical expectations~\cite{Wang:2022fbk}.
Knowing the product BFs of $D\to a_0(980)e^+\nu_e$ with $a_0(980)\to \pi\eta$, it is possible to predict the product BFs of $D\to a_0(980)e^+\nu_e$ with $a_0(980)\to K\bar K$ according to $\mathcal B(a_0(980) \to K \bar K)/\mathcal B(a_0(980) \to \eta \pi)$ = $0.172\pm0.019$~\cite{ref::pdg2022}, and outlined in Table~\ref{bfs}.
Figure~\ref{fig:fey} shows the tree-level Feynman diagrams of $D^0\to K_{S}^{0}K^{-}e^{+}\nu_{e}$, $D^+\to K_{S}^{0}K_{S}^{0}e^{+}\nu_{e}$, and $D^+\to K^{+}K^{-}e^{+}\nu_{e}$.
These decay processes can be reconstructed using charged tracks alone, providing a cleaner environment for studying the $a_0(980)$ meson~\cite{BESIII:2018sjg}.

\begin{table*}[htp]
\renewcommand{\arraystretch}{1.4}
\centering
\caption{The measured product BFs of $D\to a_0(980)e^+\nu_e$ with $a_0(980) \to \eta \pi$ and the expected  BFs of $D\to a_0(980)e^+\nu_e$ with $a_0(980) \to K\bar K$.}
\begin{tabular}{ccccc}
\hline
\hline
 & $a_0(980)[\to \pi\eta]e^+\nu_e$~\cite{BESIII:2018sjg}&$a_0(980)[\to K_S^0K^-]e^+\nu_e$&$a_0(980)[\to K_S^0K_S^0]e^+\nu_e$ &$a_0(980)[\to K^+K^-]e^+\nu_e$ \\
 BFs&($\times 10^{-4}$)&($\times 10^{-5}$)&($\times 10^{-5}$)&($\times 10^{-5}$)\\ \hline
 $D^{0}$& $1.33^{+0.33}_{-0.29}\pm0.09$ &  $1.14^{+0.28}_{-0.24}\pm0.07$ &-& -\\
 $D^{+}$& $1.66^{+0.81}_{-0.66}\pm0.11$ &-&$0.71^{+0.34}_{-0.28}\pm0.05$& $2.86^{+1.39}_{-1.13}\pm0.19$\\
\hline
\hline
\end{tabular}
\label{bfs}
\end{table*}

In this paper, we present the first searches for the semileptonic decays $D^0\to K_{S}^{0}K^{-}e^{+}\nu_{e}$, $D^+\to K_{S}^{0}K_{S}^{0}e^{+}\nu_{e}$, and $D^+\to K^{+}K^{-}e^{+}\nu_{e}$.
This analysis is based on data samples collected by the BESIII detector at a center-of-mass energy of $\sqrt s =$ 3.773 GeV in 2010, 2011, and 2021, corresponding to a total integrated luminosity of 7.9 fb$^{-1}$~\cite{Ablikim:2013ntc}.
Throughout this paper, charge-conjugate channels are always implied.

\section{BESIII detector and Monte Carlo simulation}
The BESIII detector~\cite{Ablikim:2009aa} records symmetric $e^+e^-$ collisions
provided by the BEPCII storage ring~\cite{Yu:IPAC2016-TUYA01}
in the center-of-mass energy range from 2.00 to 4.95~GeV,
with a peak luminosity of $1 \times 10^{33}\;\text{cm}^{-2}\text{s}^{-1}$
achieved at $\sqrt{s} = 3.77\;\text{GeV}$.
BESIII has collected large data samples in this energy region~\cite{Ablikim:2019hff,EcmsMea,EventFilter}. The cylindrical core of the BESIII detector~\cite{detvis} covers 93\% of the full solid angle and consists of a helium-based
 multilayer drift chamber~(MDC), a plastic scintillator time-of-flight
system~(TOF), and a CsI(Tl) electromagnetic calorimeter~(EMC),
which are all enclosed in a superconducting solenoidal magnet
providing a 1.0~T magnetic field.

The solenoid is supported by an
octagonal flux-return yoke with resistive plate counter muon
identification modules interleaved with steel.
The charged-particle momentum resolution at $1~{\rm GeV}/c$ is
$0.5\%$, and the
$dE/ dx$
resolution is $6\%$ for electrons
from Bhabha scattering. The EMC measures photon energies with a
resolution of $2.5\%$ ($5\%$) at $1$~GeV in the barrel (end-cap)
region. The time resolution in the TOF barrel region is 68~ps, while
that in the end-cap region was 110~ps.
The end cap TOF system was upgraded in 2015 using multigap resistive plate chamber
technology, providing a time resolution of 60~ps, which benefits 83\% of the data used in this analysis~\cite{etof}.

Simulated event samples produced with the {\sc geant4}-based~\cite{geant4} Monte Carlo (MC) package which
includes the geometric description of the BESIII detector and the
detector response, are used to determine the detection efficiency
and estimate the backgrounds. The simulation includes the beam-energy spread and initial-state radiation (ISR) in the $e^+e^-$
annihilations modeled with the generator {\sc kkmc}~\cite{kkmc}.
The inclusive MC sample includes the production of $D\bar{D}$
pairs (including quantum coherence for the neutral $D$ channels), the non-$D\bar{D}$ decays of the $\psi(3770)$, the ISR
production of the $J/\psi$ and $\psi(3686)$ states, and the
continuum processes incorporated in {\sc kkmc}~\cite{kkmc}.
All particle decays are modelled with {\sc
evtgen}~\cite{evtgen} using the BFs either taken from the
Particle Data Group~\cite{ref::pdg2022}, when available, or otherwise estimated with {\sc lundcharm}~\cite{lundcharm}. Final-state radiation (FSR) from charged particles is incorporated with the {\sc
photos} package~\cite{photos}. In this paper, the inclusive MC samples are used to determine the selection efficiencies and estimate the backgrounds.

The semileptonic decays $D^{0}\to K_{S}^{0} K^{-} e^{+}\nu_{e}$, $D^{+}\to K_{S}^{0} K_{S}^{0} e^{+}\nu_{e}$, and $D^{+}\to K^{+}K^{-} e^{+}\nu_{e}$ are simulated with a DIY generator with the Flatt\'{e} formula to describe the $a_0(980)$ resonance.
The Flatt\'{e} formula takes into account the mass, width, and coupling constants of the resonance to calculate its contribution to the decay rates. The mass of the $a_0(980)$ resonance is fixed at
0.990~GeV/$c^2$, while the two coupling constants coupled to $\eta\pi$~(g1) and $K\bar K$~(g2) are fixed at 0.341~(GeV/$c^2$)$^2$ and 0.304~(GeV/$c^2$)$^2$, respectively, as determined in Refs.~\cite{Bugg:1994mg,Bugg:2008ig,BESIII:2016tqo}.

\section{Method}

At $\sqrt s=3.773$\rm \,GeV, the $D^0\bar D^0$ or $D^+D^-$ meson pairs are produced from $\psi(3770)$ decays
without accompanying hadrons, which provides an ideal opportunity to study semileptonic decays of $D$ mesons using the double-tag~(DT) method~\cite{mark3}.
In the first step of the analysis, the single-tag (ST) $\bar D^0$ mesons are reconstructed via the hadronic-decay modes of
$\bar D^0\to K^+\pi^-$, $K^+\pi^-\pi^0$, and $K^+\pi^-\pi^-\pi^+$;
while the ST $D^-$ mesons are reconstructed via the decays
$D^-\to K^{+}\pi^{-}\pi^{-}$, $K^0_{S}\pi^{-}$, $K^{+}\pi^{-}\pi^{-}\pi^{0}$, $K^0_{S}\pi^{-}\pi^{0}$,
$K^0_{S}\pi^{+}\pi^{-}\pi^{-}$, and $K^{+}K^{-}\pi^{-}$.
Then the semileptonic decays of $D$ meson candidates are reconstructed with the remaining tracks which have not been used in the ST selection.
The event, in which the semileptonic decays $D^0\to K_{S}^{0}K^{-}e^{+}\nu_{e}$, $D^+\to K_{S}^{0}K_{S}^{0}e^{+}\nu_{e}$, and $D^+\to K^{+}K^{-}e^{+}\nu_{e}$ are reconstructed in the systems recoiling against the ST $\bar D$ mesons, is called a DT event.
The product BFs of $D^0\to K_{S}^{0}K^{-}e^{+}\nu_{e}$, $D^+\to K_{S}^{0}K_{S}^{0}e^{+}\nu_{e}$, and $D^+\to K^{+}K^{-}e^{+}\nu_{e}$ are determined by
\begin{equation}
    \label{br}
   {
{\mathcal B}_{\rm SL} = \frac{N_{\rm DT}}{N^{\rm tot}_{\rm ST} \bar \epsilon_{\rm sig} ({\mathcal B}_{K_S^{0}})^k},}
    \end{equation}
where $N_{\rm ST}^{\rm tot}$ and $N_{\rm DT}$ are the yields of the ST $\bar D^0(D^-)$ mesons and the DT signal events in data, respectively; $\mathcal B_{K^0_S}$ is the BF of $K^0_S\to \pi^+\pi^-$ quoted from the Particle Data Group~\cite{ref::pdg2022}; $k$ is the number of $K_{S}^{0}$ mesons in the final state of DT side, and
$\bar \epsilon_{\rm sig}$ is the average signal efficiency weighted by the measured yields of tag modes $i$ in the data, i.e.,
   \begin{equation}
    \label{br}
   {
\bar{{\mathcal \epsilon}}_{\rm sig} = \frac{\sum_i (N^i_{\rm ST}\cdot \epsilon^i_{\rm DT}/\epsilon^i_{\rm ST})}{N^{\rm tot}_{\rm ST}},}
    \end{equation}
where $N^i_{\rm ST}$ are the yields of the ST candidates observed in data, $\epsilon^i_{\rm ST}$ is the efficiency of reconstructing the ST mode $i$ (referred to as the ST efficiency),
and $\epsilon^i_{ \rm DT}$ is the efficiency of finding the ST mode $i$ and the $D^0\to K_{S}^{0}K^{-}e^{+}\nu_{e}$, $D^+\to K_{S}^{0}K_{S}^{0}e^{+}\nu_{e}$, and $D^+\to K^{+}K^{-}e^{+}\nu_{e}$ decay simultaneously (referred to as the DT efficiency).

\section{Single Tag selection}

Charged tracks detected in the MDC (except for those used for $K^0_S$ reconstruction) are required to be within a polar angle ($\theta$) range of $|\rm{cos\theta}|<0.93$, where $\theta$ is defined with respect to the $z$-axis,
which is the symmetry axis of the MDC.
The distance of closest approach to the interaction point (IP) must be less than 10\,cm
along the $z$-axis, $|V_{z}|$,  and less than 1\,cm in the transverse plane, $|V_{xy}|$.
Simple particle identification~(PID) for charged tracks combines measurements of the specific ionization energy loss in the MDC~(d$E$/d$x$) and the flight time in the TOF to form likelihoods $\mathcal{L}(h)~(h=p,K,\pi)$ for each hadron $h$ hypothesis.
Charged kaons and pions are identified by comparing the likelihoods for the kaon and pion hypotheses, $\mathcal{L}(K)>\mathcal{L}(\pi)$ and $\mathcal{L}(\pi)>\mathcal{L}(K)$, respectively.

Each $K_{S}^0$ candidate is reconstructed from two oppositely charged tracks satisfying $|V_{z}|<20$~cm.
The two charged tracks are assigned as $\pi^+\pi^-$ without imposing PID criteria.
They are constrained to originate from a common vertex, requiring an invariant mass within $(0.487,0.511)$~GeV/$c^2$.
The decay length of the $K^0_S$ candidate is required to be greater than
twice the vertex resolution away from the IP.
The quality of the vertex fits (primary-vertex fit and secondary-vertex fit) is ensured by a
requirement on the $\chi^2$ ($\chi^2 < 100$).

Photon candidates are identified using showers in the EMC.
The deposited energy of each shower must be more than 25~MeV in the barrel region ($|\!\cos \theta|< 0.80$) and more than 50~MeV in the end-cap region ($0.86 <|\!\cos \theta|< 0.92$).
Showers are required to be separated from other charged tracks by an angle greater than $10^\circ$
 in order to eliminate activity induced by tracks.
To suppress electronic noise and showers unrelated to the event, the difference between the EMC time and the event start time is required to be within
[0, 700]\,ns.
For $\pi^0$ candidates, the invariant mass of the photon pair is required to be within $(0.115,\,0.150)$\,GeV$/c^{2}$.
To improve the resolution, a kinematic fit is performed, where the diphoton invariant mass is constrained to the known $\pi^{0}$ mass~\cite{ref::pdg2022}. The momenta obtained from the kinematic fit are used in the subsequent analysis.

In the selection of $\bar D^0\to K^+\pi^-$ events, the backgrounds from cosmic rays and Bhabha events are rejected by using the same requirements described in Ref.~\cite{deltakpi}.
The two charged tracks must have a TOF time difference of less than 5~ns. They must not be consistent with being a muon or electron-positron pair. Additionally, there must be at least one EMC shower with energy deposited larger than 50~MeV, or at least one additional charged track detected in the MDC.

To separate the ST $\bar D$ mesons from combinatorial backgrounds, we define the energy difference $\Delta E\equiv E_{\bar D}-E_{\mathrm{beam}}$ and the beam-constrained mass $M_{\rm BC}\equiv\sqrt{E_{\mathrm{beam}}^{2}/c^{4}-|\vec{p}_{\bar D}|^{2}/c^{2}}$, where $E_{\mathrm{beam}}$ is the beam energy, and $E_{\bar D}$ and $\vec{p}_{\bar D}$ are the total energy and momentum of the $\bar D$ candidate in the $e^+e^-$ center-of-mass frame, respectively.
If there is more than one $\bar D$ candidate in a given ST mode, that candidate with the least $|\Delta E|$ is kept for the subsequent analysis.
The $\Delta E$ requirements and ST efficiencies are listed in Table~\ref{ST:realdata}.

\begin{figure}[htbp]\centering
\includegraphics[width=1.0\linewidth]{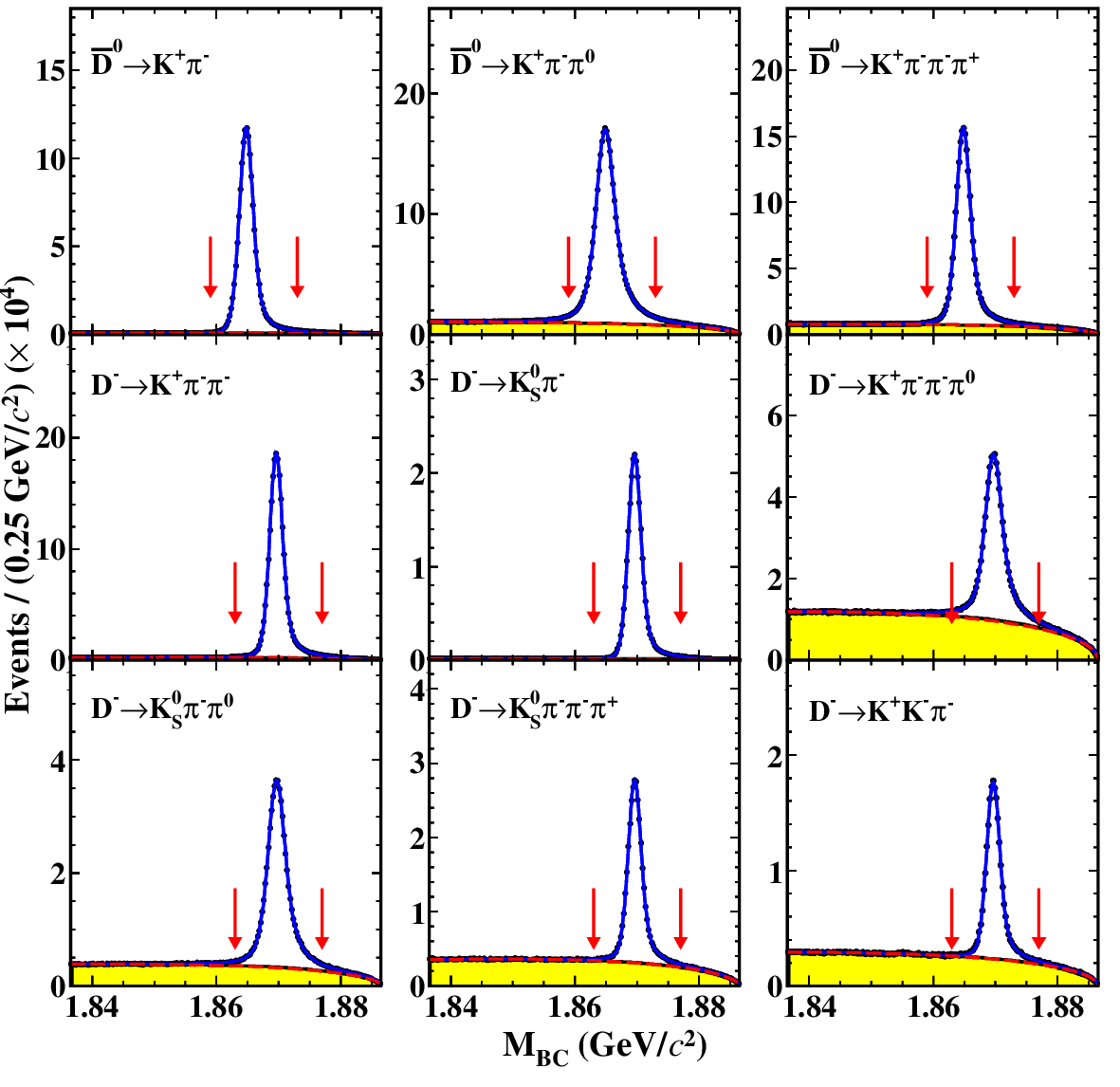}
\caption{
Fits to the $M_{\rm BC}$ distributions of the ST $\bar D$ candidates.
In each plot, the points with error bars correspond to the data, the blue curves are the best fits, and the red dashed curves describe the fitted combinatorial background shapes. The yellow normalization histograms show the simulated background contributions from the inclusive MC sample. The pair of red arrows indicate the $M_{\rm BC}$ signal window.}\label{fig:datafit_Massbc}
\end{figure}

The ST yields are extracted by performing unbinned maximum likelihood fits to the corresponding $M_{\rm BC}$ distribution. In the fit, the signal shape is derived from the MC-simulated signal shape convolved with a double-Gaussian function to compensate for the resolution difference between the data and the MC simulation.
The background shape is described by the ARGUS function~\cite{argus}, with the endpoint parameter fixed at 1.8865~GeV/$c^{2}$ corresponding to $E_{\rm beam}$.
Figure~\ref{fig:datafit_Massbc} shows the fits to the $M_{\rm BC}$ distributions of the accepted ST candidates in data for different ST modes. The candidates with $M_{\rm BC}$ within $(1.859,1.873)$ GeV/$c^2$ for $\bar D^0$ tags and $(1.863,1.877)$ GeV/$c^2$ for $D^-$ tags are kept for further analyses. Summing over the tag modes gives the total yields of ST $\bar D^0$ and $D^-$ mesons to be $(6306.8 \pm 2.8_{\rm stat})\times 10^3$ and $(4149.9\pm2.3_{\rm stat})\times 10^3$, respectively.

\begin{table}
\renewcommand{\arraystretch}{1.2}
\centering
\caption {The $\Delta E$ requirements, the measured ST $\bar D$ yields in the data, and the ST efficiencies ($\epsilon_{\rm ST}^{i}$) for nine tag modes. The uncertainties are statistical only.}
\scalebox{0.87}{
\begin{tabular}{cccc}
\hline
\hline
Tag mode & $\Delta E$(GeV)  &  $N^{i}_{\rm data}(\times 10^3)$  &  $\epsilon^i_{\rm ST}(\%)    $       \\\hline
$\bar D^0\to K^+\pi^-$                    &  $(-0.027,0.027)$ & $1449.5\pm1.2$&$64.95\pm0.01$\\
$\bar D^0\to K^+\pi^-\pi^0$             &  $(-0.062,0.049)$ & $2913.1\pm2.0$&$35.52\pm0.00$\\
$\bar D^0\to K^+\pi^-\pi^-\pi^+$       &  $(-0.026,0.024)$ & $1944.1\pm1.5$&$40.42\pm0.01$\\
\hline
$D^-\to K^+\pi^-\pi^-$                   &  $(-0.025,0.024)$ & $2164.0\pm1.5$&$51.17\pm0.01$\\
$D^-\to K^{0}_{S}\pi^{-}$                 &  $(-0.025,0.026)$ & $250.4\pm0.5$&$50.63\pm0.02$\\
$D^-\to K^{+}\pi^{-}\pi^{-}\pi^{0}$     &  $(-0.057,0.046)$ & $689.0\pm1.1$&$25.50\pm0.01$\\
$D^-\to K^{0}_{S}\pi^{-}\pi^{0}$         &  $(-0.062,0.049)$ & $558.4\pm0.9$&$26.28\pm0.01$\\
$D^-\to K^{0}_{S}\pi^{-}\pi^{-}\pi^{+}$ &  $(-0.028,0.027)$ & $300.5\pm0.6$&$28.97\pm0.01$\\
$D^-\to K^{+}K^{-}\pi^{-}$                &  $(-0.024,0.023)$ & $187.3\pm0.5$&$41.06\pm0.02$\\
\hline
\hline
          \end{tabular}
          }
          \label{ST:realdata}
          \end{table}

\section{Double Tag selection}

The candidates for $D^0\to K_{S}^{0}K^{-}e^{+}\nu_{e}$, $D^+\to K_{S}^{0}K_{S}^{0}e^{+}\nu_{e}$, and $D^+\to K^{+}K^{-}e^{+}\nu_{e}$ are selected from the remaining tracks in the presence of the tagged $\bar D$ candidates.
We require that there are four, five, and three charged tracks~($N^{\rm charge}_{\rm extra}$) reconstructed in the $D^0\to K_{S}^{0}K^{-}e^{+}\nu_{e}$, $D^+\to K_{S}^{0}K_{S}^{0}e^{+}\nu_{e}$ and $D^+\to K^{+}K^{-}e^{+}\nu_{e}$ modes, respectively.

Candidates for $K^\pm$ and $K_{S}^{0}$ are selected with the same criteria as those used in the ST selection. The positron is identified using the measured information in the MDC, TOF and EMC. The combined likelihoods ($\mathcal{L}'$)
under the positron, pion, and kaon hypotheses are obtained.
Positron candidates are required to satisfy $\mathcal{L}'(e)>0.001$ and $\mathcal{L}'(e)/(\mathcal{L}'(e)+\mathcal{L}'(\pi)+\mathcal{L}'(K))>0.8$. To reduce background
from hadrons and muons, the positron candidate is further required to have a deposited energy in the EMC greater than 0.8 times its momentum obtained in the MDC.

To suppress the backgrounds containing extra $\pi^0$ mesons, we require that there is no additional combinations of two photons~($N_{\rm extra \pi^0}$) that satisfy the requirements for a $\pi^0$ meson in the event selection. To reject contamination from the hadronic decays involving a $\pi^0$, e.g., $D^0 \to K_S^0 K^- \pi^+ \pi^0$, $D^+ \to K_S^0 K_S^0 \pi^+ \pi^0$, $D^+ \to K^+ K^- \pi^+ \pi^0$, the maximum energies of any extra photons ($E_{\text{extra~}\gamma}^{\rm max}$) which have not been used in the event selection are required to be less than  0.20 GeV, 0.17 GeV, and 0.25 GeV for $D^0\to K_{S}^{0}K^{-}e^{+}\nu_{e}$, $D^+\to K_{S}^{0} K_{S}^{0}e^{+}\nu_{e}$, and $D^+\to K^{+} K^{-}e^{+}\nu_{e}$, respectively.
To suppress the backgrounds from the hadronic decays $D^0\to K^-\pi^+\pi^+\pi^-$, $D^+\to 2(\pi^+\pi^-)\pi^+$, and $D^+\to K^+K^-\pi^+$,
the invariant masses of the $K\bar{K}e$ combinations are required to be less than 1.74 GeV/$c^2$, 1.77 GeV/$c^2$, and 1.75 GeV/$c^2$ for $D^0\to K_{S}^{0}K^{-}e^{+}\nu_{e}$, $D^+\to K_{S}^{0} K_{S}^{0}e^{+}\nu_{e}$, and $D^+\to K^{+} K^{-}e^{+}\nu_{e}$, respectively.
An additional requirement is deployed in the selection of $D^+\to K^{+}K^{-}e^{+}\nu_{e}$ events, to suppress the background from $D^+\to {K}^{+}K^{-}\pi^{+}\pi^{0}$ decays due to misidentifying a pion as an electron: the opening angle between the missing momentum and the most energetic shower, $\theta_{\vec{p}_{\mathrm{miss}},\gamma}$, is required to satisfy $\cos\theta_{\vec{p}_{\mathrm{miss}},\gamma}<0.86$. These requirements have been optimized according to the Punzi metric~\cite{punzi}.

Events containing neutrinos cannot be fully reconstructed.
To select semileptonic signal candidates, we define $U_{\mathrm{miss}}\equiv E_{\mathrm{miss}}-|\vec{p}_{\mathrm{miss}}|c$, where $E_{\mathrm{miss}}$ and $\vec{p}_{\mathrm{miss}}$
are the missing energy and momentum of the DT event in the $e^+e^-$ center-of-mass frame, respectively.
These quantities are calculated by $E_{\mathrm{miss}}\equiv E_{\mathrm{beam}}-E_{K_S^0\,(K_S^0)\,(K^+)}-E_{K^-\,(K_S^{0})\,(K^-)}-E_{e^{+}}$ and $\vec{p}_{\mathrm{miss}}\equiv\vec{p}_{D}-\vec{p}_{K_S^0\,(K_S^0)\,(K^+)}-\vec{p}_{K^-\,(K_S^{0})\,(K^-)}-\vec{p}_{e^{+}}$, where $E_{K_S^{0}\,(K^+)\,(K^-)\,(e^+)}$ and $\vec{p}_{K_S^{0}\,(K^+)\,(K^-)\,(e^+)}$ are the measured energy and momentum of the $K_S^{0}\,(K^+)\,(K^-)\,(e^+)$ candidates, respectively, and $\vec{p}_{D}\equiv-\hat{p}_{\bar D} \sqrt{E_{\mathrm{beam}}^{2}/c^{2}-m_{\bar D}^{2} c^{2} }$, where
$\hat{p}_{\bar D}$ is the unit vector in the momentum direction of the ST $\bar D$ meson and $m_{\bar D}$ is the known $\bar D$ mass~\cite{ref::pdg2022}.
For the decays $D^0\to K_{S}^{0}K^{-}e^{+}\nu_{e}$ and $D^+\to K^{+}K^{-}e^{+}\nu_{e}$, the backgrounds from $D^0\to K_{S}^{0}\pi^{-}e^{+}\nu_{e}$ or $D^+\to \pi^{+}K^{-}e^{+}\nu_{e}$ due to the misidentification of the kaon are suppressed with the requirement of $0.16<U_{\rm miss}^{\pi}<0.31$ GeV and $0.17<U_{\rm miss}^{\pi}<0.32$ GeV, where  $E_{\rm miss}^{\pi}$ and $U_{\rm miss}^{\pi}$ are calculated by replacing the $K$ mass with the $\pi$ mass in the previously defined quantities.
Here, the beam energy and the nominal $\bar D$ mass are used to improve the $U_{\mathrm{miss}}$ resolution.

The average signal efficiencies in the presence of the ST $\bar D$ mesons are $(11.06\pm0.07)\%$, $(8.51\pm0.06)\%$, and $(13.06\pm0.07)\%$ for $D^0\to K_{S}^{0}K^{-}e^{+}\nu_{e}$, $D^+\to K_{S}^{0}K_{S}^{0}e^{+}\nu_{e}$ and $D^+\to K^{+}K^{-}e^{+}\nu_{e}$, respectively. These efficiencies do not include the BF of $K^0_S\to \pi^+\pi^-$.

\section{Results}

\begin{figure*}[htbp]
\includegraphics[width=1.0\linewidth]{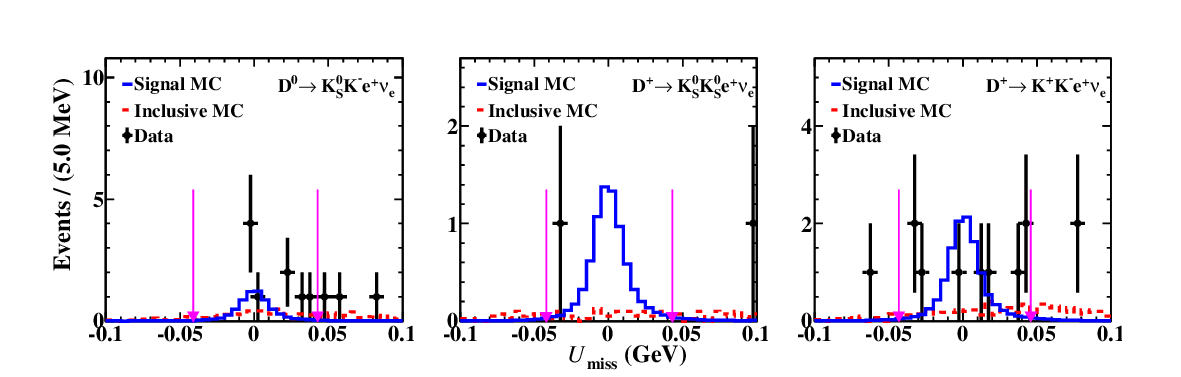}
\caption{The $U_{\rm miss}$ distributions of the accepted candidate events for $D^0\to K_{S}^{0} K^{-}e^+\nu_e$, $D^+\to K_{S}^{0}K_{S}^{0}e^{+}\nu_{e}$ and $D^+\to K^{+} K^{-}e^{+}\nu_{e}$. The dots with error bars
are data, the blue histograms are the signal MC samples normalized with a product BF of $2\times 10^{-4}$ and the red dashed lines are the inclusive MC sample.
The regions inside the pair of magenta arrows denote the signal regions.
 }
\label{fig:data}
\end{figure*}

Figure~\ref{fig:data} shows the  $U_{\rm miss}$ distributions of the candidate events for $D^0\to K_{S}^{0}K^{-}e^{+}\nu_{e}$, $D^+\to K_{S}^{0}K_{S}^{0}e^{+}\nu_{e}$ and $D^+\to K^{+}K^{-}e^{+}\nu_{e}$ selected from data.
The signal yields are obtained by counting the events in the $U_{\rm miss}$ signal regions.
Based on the MC study, the signal regions are defined as
$[-0.041, 0.043]$~GeV, $[-0.042, 0.043]$~GeV, and $[-0.043, 0.046]$~GeV for $D^0\to K_{S}^{0}K^{-}e^{+}\nu_{e}$, $D^+\to K_{S}^{0}K_{S}^{0}e^{+}\nu_{e}$ and $D^+\to K^{+}K^{-}e^{+}\nu_{e}$, respectively, which correspond to intervals that are three times the resolution of the signal peaks.
The yields in the signal regions ($N^{\rm sig}$) of the candidates for $D^0\to K_{S}^{0}K^{-}e^{+}\nu_{e}$, $D^+\to K_{S}^{0}K_{S}^{0}e^{+}\nu_{e}$, and $D^+\to K^{+}K^{-}e^{+}\nu_{e}$ are determined to be 9, 1, and 9, respectively.
Based on the inclusive MC sample, the background yields ($N^{\rm bkg}$) are estimated to be 4.5, 1.1, and 3.5 for $D^0\to K_{S}^{0} K^{-}e^+\nu_e$, $D^+\to K_{S}^{0}K_{S}^{0}e^{+}\nu_{e}$, and $D^+\to K^{+} K^{-}e^{+}\nu_{e}$, respectively.

Since no significant excesses are observed above background, we set the upper limits on the BFs of $D^0\to K_{S}^{0} K^{-}e^+\nu_e$, $D^+\to K_{S}^{0} K_{S}^{0}e^{+}\nu_{e}$, and $D^+\to K^{+} K^{-}e^{+}\nu_{e}$. Upper limits on the numbers of signal events at the 90\% confidence level (C.L.) are calculated by using a frequentist method~\cite{trokle} with an unbound profile likelihood treatment of systematic uncertainties (see below), as implemented by the $\textsc{TRolke}$ package in the ROOT software~\cite{ROOT}
with the quantities of $N^{\rm sig}$, $N^{\rm bkg}$, $\bar \epsilon_{\rm sig}$, and the total systematic uncertainty ($\delta_{\rm syst}$) as input. Here, the numbers of the signal and background events
are assumed to follow a Poisson distribution, while the detection efficiency is assumed to follow a Gaussian distribution.
Finally, the upper limits on the BFs of $D^0\to K_{S}^{0} K^{-}e^+\nu_e$, $D^+\to K_{S}^{0} K_{S}^{0}e^{+}\nu_{e}$, and $D^+\to K^{+} K^{-}e^{+}\nu_{e}$ at the 90\% C.L. are set to be
$2.13\times10^{-5}$, $1.54\times10^{-5}$ and $2.10\times10^{-5}$, respectively.

\section{Systematic uncertainty}

With the DT method, many systematic uncertainties associated with the ST selection cancel and do not affect the BF measurement.

The uncertainty associated with the ST yield $N_{\rm ST}^{\rm tot}$, is assigned as 0.1\% after varying the signal, background shapes and floating the parameters of one Gaussian in the fit.
The tracking and PID efficiencies of $e^\pm$ are studied with a control sample of radiative Bhabha events, and those of the $K^\pm$ are studied by analyzing DT $D^{0}\bar{D}^0(D^+D^-)$ events, where the control samples comprise hadronic decays of $D^{0}\to K^-\pi^+$, $D^{0}\to K^- \pi^+\pi^0$, $D^{0}\to K^-\pi^+\pi^+\pi^-$ versus $D^0 \to K^+\pi^-$, $D^{0}\to K^+ \pi^-\pi^0$, $D^{0}\to K^+\pi^-\pi^-\pi^+$ as well as $D^+ \to K^-\pi^+\pi^+$ versus $D^+ \to K^+\pi^-\pi^-$.
The systematic uncertainty due to tracking efficiencies is assigned as 1.0\% for both $K^\pm$ and $e^\pm$;
the systematic uncertainty due to PID is assigned as 1.0\% for $K^\pm$ and $e^\pm$.
The uncertainty from the $K_{S}^{0}$ reconstruction is 1.5\%, which is obtained by studying control samples of $J/\psi\to K^*(892)^\pm K^\mp$ and $J/\psi \to \phi K^0_SK^\pm\pi^\mp$~\cite{ks} decays.

The uncertainty in the BF of $K_{S}^{0}\to \pi^+\pi^-$ is 0.1\%~\cite{ref::pdg2022}.
The uncertainties due to the limited size of MC samples are 0.6\%, 0.7\%, and 0.5\% for
$D^0\to K_{S}^{0} K^{-}e^+\nu_e$, $D^+\to K_{S}^{0} K_{S}^{0}e^{+}\nu_{e}$ and $D^+\to K^{+} K^{-}e^{+}\nu_{e}$, respectively.

The signal MC samples in this study are generated with the DIY generator. The $M_{K\bar{K}}$ propagator is parameterized with a Flatt\'{e} formula~\cite{BESIII:2016tqo}.
To estimate the uncertainty from the MC generator, the quoted coupling constants ($g_{1}$ and $g_{2}$) are varied by $\pm 1\sigma$ to produce alternative signal MC samples.
The maximum changes in the DT efficiency between the DIY MC and the alternative signal MC samples are assigned to be 2.7\%, 1.1\%, and 0.9\% for $D^0\to K_{S}^{0} K^{-}e^+\nu_e$, $D^+\to K_{S}^{0} K_{S}^{0}e^{+}\nu_{e}$, and $D^+\to K^{+} K^{-}e^{+}\nu_{e}$, respectively.

The combined systematic uncertainties from the $E^{\rm max}_{\rm extra\gamma}$, $N_{\rm extra\pi^{0}}$ and $ N^{\rm charge}_{\rm extra}$ requirements are estimated to be 1.6\%, 2.0\%, and 0.9\% for $D^0\to K_{S}^{0} K^{-}e^+\nu_e$, $D^+\to K_{S}^{0} K_{S}^{0}e^{+}\nu_{e}$ and $D^+\to K^{+} K^{-}e^{+}\nu_{e}$, respectively, which are assigned using from studies of  DT samples of $D^0\to K^-e^+\nu_e$ and $D^+\to K^0_Se^+\nu_e$ reconstructed versus the same ST modes used in the baseline analysis.

The uncertainties from the $M_{K\bar{K}e}$, $U_{\rm miss}^{\pi \to K}$, and $\cos\theta_{\vec{p}_{\mathrm{miss}},\gamma}$ requirements are obtained by varying their values by
$\pm 10$~MeV/$c^2$, $\pm 1$~MeV, $\pm 0.01$,
respectively, following the method defined in Refs.~\cite{BESIII:2017pez,BESIII:2018nzb,BESIII:2021uqr}. The maximum changes of the BF upper limits are taken as the associated systematic uncertainties.

Due to the limited sampe size, only the resonant $K\bar{K}$ contributions in $D^0\to K_{S}^{0} K^{-}e^+\nu_e$, $D^+\to K_{S}^{0} K_{S}^{0}e^{+}\nu_{e}$, and $D^+\to K^{+} K^{-}e^{+}\nu_{e}$ are considered. The associated systematic uncertainty is assigned by using the alternative signal MC samples, mixed with 20\% of non-resonant $D\to K_S^0K~(K_S^0K_S^0)~(KK)e^{+}\nu_e$ and 80\% of $D\to a_0(980)(\to K_S^0K~(K_S^0K_S^0)~(KK))e^{+}\nu_e$ decays. This is a conservative estimation as the largest known non-resonant contribution in the charm sector is only about 6.0\% in the $D^+\to K^-\pi^+e^+\nu_e$ decay ~\cite{BESIII:2015hty}.
The differences between the nominal and alternative signal efficiencies, 5.1\%, 5.0\% and 4.6\%, are taken as the systematic uncertainties for the BFs of the decays $D^0\to K^0_SK^-e^+\nu_e$, $D^+\to K^0_SK^0_Se^+\nu_e$, and $D^+\to K^+K^-e^+\nu_e$, respectively. The uncertainties due to the BFs of the $D^0$ and $D^+$ decays and the cross sections of $D^0\bar D^0$ and $D^+D^-$ are negligible.

The total systematic uncertainty is obtained by adding the individual components in quadrature, assuming that all sources are uncorrelated.
Table~\ref{sys} summarizes the sources of the systematic uncertainties in the BF measurements.

\begin{table*}[htp]
      \centering
       \caption{Relative systematic uncertainties ($\delta_{\rm syst}$, in \%) in the BF measurements.}
      \begin{tabular}{cccc}
    \hline   \hline
   Source &$D^0\to K_{S}^{0} K^{-}e^+\nu_e$&$D^+\to K_{S}^{0} K_{S}^{0}e^{+}\nu_{e}$ & $D^+\to K^{+} K^{-}e^{+}\nu_{e}$ \\
    \hline
 $N_{\rm ST}^{\rm tot}$                                    &0.1 &0.1&0.1 \\
        $K$/$e$ tracking                                   &2.0 &1.0&3.0  \\
        $K$/$e$ PID                                        &2.0 &1.0&3.0   \\
        $K_{S}^0$ reconstruction                           &1.5 &3.0&...  \\
        Quoted $\mathcal B$                                &0.1 &0.2&... \\
        MC sample size                                      &0.6 &0.7&0.5  \\
        MC generator                                       &2.7 &1.1&0.9 \\
        $E^{\rm max}_{\rm extra\gamma} , N_{\rm extra\pi^{0}}$ and $ N^{\rm charge}_{\rm extra}$ requirements            &1.6&2.0&0.9\\
        $M_{K\bar{K}e^+}$ requirement                       &0.4&1.2&2.4  \\
        $U_{\rm miss}^{\pi}$ requirement              &1.0&...&0.5  \\
        $\cos\theta_{\vec{p}_{\mathrm{miss}},\gamma}$ requirement               &...&...&2.0 \\
        Non-resonant $K\bar{K}e\nu_{e}$ component         &5.1 &5.0&4.6 \\
    \hline
     Total                               &6.9&6.6&7.1\\
          \hline  \hline
            \end{tabular}
             \label{sys}
            \end{table*}

\section{Summary}

By analyzing 7.9 $\rm fb^{-1}$  of $e^+e^-$ annihilation data taken at $\sqrt{s}=$ 3.773 GeV, we search for the semileptonic decays $D^0\to K_{S}^{0} K^{-}e^+\nu_e$, $D^+\to K_{S}^{0} K_{S}^{0}e^{+}\nu_{e}$, and $D^+\to K^{+} K^{-}e^{+}\nu_{e}$. No significant signals are observed.
The upper limits on the BFs of $D^0\to K_{S}^{0} K^{-}e^+\nu_e$, $D^+\to K_{S}^{0} K_{S}^{0}e^{+}\nu_{e}$, and $D^+\to K^{+} K^{-}e^{+}\nu_{e}$ are set to be
$2.13\times10^{-5}$, $1.54\times10^{-5}$, and $2.10\times10^{-5}$ at the 90\% C.L., respectively. These upper limits are comparable to the expected product BFs of the individual decays. An increased data set corresponding to an integrated luminosity of 20 $\rm fb^{-1}$  taken at $\sqrt{s}=$ 3.773 GeV at BESIII will be available in the near future~\cite{Ablikim:2019hff,Ke:2023qzc,Li:2021iwf}.
This larger sample will offer an opportunity to further improve the sensitivity of the search for these semileptonic decays.

\section{Acknowledgement}
The BESIII Collaboration thanks the staff of BEPCII and the IHEP computing center for their strong support. This work is supported in part by National Key R\&D Program of China under Contracts Nos. 2020YFA0406400, 2020YFA0406300; National Natural Science Foundation of China (NSFC) under Contracts Nos. 11635010, 11735014, 11835012, 11935015, 11935016, 11935018, 11961141012, 12025502, 12035009, 12035013, 12061131003, 12192260, 12192261, 12192262, 12192263, 12192264, 12192265, 12221005, 12225509, 12235017; the Chinese Academy of Sciences (CAS) Large-Scale Scientific Facility Program; the CAS Center for Excellence in Particle Physics (CCEPP); Joint Large-Scale Scientific Facility Funds of the NSFC and CAS under Contract No. U1832207; CAS Key Research Program of Frontier Sciences under Contracts Nos. QYZDJ-SSW-SLH003, QYZDJ-SSW-SLH040; 100 Talents Program of CAS; The Institute of Nuclear and Particle Physics (INPAC) and Shanghai Key Laboratory for Particle Physics and Cosmology; European Union's Horizon 2020 research and innovation programme under Marie Sklodowska-Curie grant agreement under Contract No. 894790; German Research Foundation DFG under Contracts Nos. 455635585, Collaborative Research Center CRC 1044, FOR5327, GRK 2149; Istituto Nazionale di Fisica Nucleare, Italy; Ministry of Development of Turkey under Contract No. DPT2006K-120470; National Research Foundation of Korea under Contract No. NRF-2022R1A2C1092335; National Science and Technology fund of Mongolia; National Science Research and Innovation Fund (NSRF) via the Program Management Unit for Human Resources \& Institutional Development, Research and Innovation of Thailand under Contract No. B16F640076; Polish National Science Centre under Contract No. 2019/35/O/ST2/02907; The Swedish Research Council; U. S. Department of Energy under Contract No. DE-FG02-05ER41374.

\end{document}

%% file: authorlist_2023-08-29.tex
\author{
M.~Ablikim$^{1}$, M.~N.~Achasov$^{4,b}$, P.~Adlarson$^{75}$, O.~Afedulidis$^{3}$, X.~C.~Ai$^{80}$, R.~Aliberti$^{35}$, A.~Amoroso$^{74A,74C}$, M.~R.~An$^{39}$, Q.~An$^{71,58}$, Y.~Bai$^{57}$
% ...
\\
\vspace{0.2cm}
(BESIII Collaboration)\\
\vspace{0.2cm} {\it
$^{1}$ Institute of High Energy Physics, Beijing 100049, People's Republic of China\\
% ...
$^{3}$ Bochum  Ruhr-University, D-44780 Bochum, Germany\\
$^{4}$ Budker Institute of Nuclear Physics SB RAS (BINP), Novosibirsk 630090, Russia\\
% ...
$^{35}$ Johannes Gutenberg University of Mainz, Johann-Joachim-Becher-Weg 45, D-55099 Mainz, Germany\\
% ...
$^{39}$ Liaoning Normal University, Dalian 116029, People's Republic of China\\
% ...
$^{57}$ Southeast University, Nanjing 211100, People's Republic of China\\
$^{58}$ State Key Laboratory of Particle Detection and Electronics, Beijing 100049, Hefei 230026, People's Republic of China\\
% ...
$^{71}$ University of Science and Technology of China, Hefei 230026, People's Republic of China\\
% ...
$^{74}$ University of Turin and INFN, (A)University of Turin, I-10125, Turin, Italy; (B)University of Eastern Piedmont, I-15121, Alessandria, Italy; (C)INFN, I-10125, Turin, Italy\\
$^{75}$ Uppsala University, Box 516, SE-75120 Uppsala, Sweden\\
% ...
$^{80}$ Zhengzhou University, Zhengzhou 450001, People's Republic of China\\
\vspace{0.2cm}
% ...
$^{b}$ Also at the Novosibirsk State University, Novosibirsk, 630090, Russia\\
% ...
}
}

%% file: draft_KKenu.bbl
\begin{thebibliography}{99}

\bibitem{ref::pdg2022}
R.L. Workman {\it et al.} (Particle Data Group),
\href{https://pdg.lbl.gov/index-2023.html}{Prog. Theor. Exp. Phys. {\bf 2022}, 083C01 (2022).}

\bibitem{BESIII:2018sjg}
M.~Ablikim {\it et al.} (BESIII Collaboration),
\href{https://journals.aps.org/prl/abstract/10.1103/PhysRevLett.121.081802}{Phys. Rev. Lett. {\bf 121}, 081802 (2018).}

\bibitem{Ablikim:2013ntc}
M.~Ablikim {\it et al.} (BESIII Collaboration),
\href{https://iopscience.iop.org/article/10.1088/1674-1137/37/12/123001/pdf}{Chin. Phys. C \textbf{37}, 123001 (2013);} \href{https://inspirehep.net/files/90f19f568cfbc1b1cfe154e1a3737876}{Phys. Lett. B \textbf{753}, 629-638 (2016).} These articles described the integrated luminosity measurement for data taken in 2010 and 2011. The integrated luminosity for data taken in 2022 is determined via a similar procedure.

\bibitem{Wang:2022fbk}
R.~M.~Wang, Y.~Qiao, Y.~J.~Zhang, X.~D.~Cheng and Y.~G.~Xu,
\href{https://arxiv.org/pdf/2301.00090.pdf}{[arXiv:2301.00090 [hep-ph]].}

\bibitem{Ablikim:2009aa}
M.~Ablikim {\it et al.} (BESIII Collaboration),
\href{https://arxiv.org/ftp/arxiv/papers/0911/0911.4960.pdf}{Nucl.\ Instrum.\ Meth.\ A {\bf 614}, 345 (2010).}

\bibitem{Yu:IPAC2016-TUYA01}
C.~H.~Yu {\it et al.},
\href{doi:10.18429/JACoW-IPAC2016-TUYA01}{Proceedings of IPAC2016, Busan, Korea, 2016.}

\bibitem{Ablikim:2019hff}
M.~Ablikim {\it et al.} (BESIII Collaboration),
\href{https://doi.org/10.1088/1674-1137/44/4/040001}{Chin. Phys. C {\bf 44}, 040001 (2020).}

\bibitem{EcmsMea}
  J.~Lu, Y.~Xiao, X.~Ji,
  \href{https://link.springer.com/journal/41605/volumes-and-issues/4-3}{Radiat. Detect. Technol. Methods {\bf 4}, 337–344 (2020).}

\bibitem{EventFilter}
  J.~W.~Zhang, L.~H.~Wu, S.~S.~Sun {\it et al.},
  \href{https://www.springer.com/journal/41605}{Radiat. Detect. Technol. Methods {\bf 6}, 289–293 (2022).}

\bibitem{detvis}
K.~X.~Huang, {\it et al.},
\href{https://doi.org/10.1007/s41365-022-01133-8}{Nucl.\ Sci.\ Tech. {\bf 33}, 142 (2022).}


\bibitem{etof}
 X.~Li {\it et al.},
 \href{https://www.springer.com/journal/41605}{Radiat. Detect. Technol. Methods {\bf 1}, 13 (2017)};
 Y.~X.~Guo {\it et al.},
 \href{https://www.springer.com/journal/41605}{Radiat. Detect. Technol. Methods {\bf 1}, 15 (2017)};
 P.~Cao {\it et al.},
 \href{https://www.sciencedirect.com/science/article/pii/S0168900219314068}{ Nucl.\ Instrum.\ Meth.\ A {\bf 953}, 163053 (2020).}


\bibitem{geant4}
S. Agostinelli {\it et al.} (GEANT4 Collaboration),
\href{https://doi.org/10.1016/S0168-9002(03)01368-8}{Nucl. Instrum. Meth. A {\bf 506}, 250 (2003).}

\bibitem{kkmc}
S. Jadach, B. F. L. Ward, and Z. Was,
\href{https://linkinghub.elsevier.com/retrieve/pii/S0010465500000485}{ Comp. Phys. Commu. {\bf 130}, 260 (2000);} \href{https://journals.aps.org/prd/abstract/10.1103/PhysRevD.63.113009}{Phys. Rev. D {\bf 63}, 113009 (2001).}

\bibitem{evtgen}
D.~J.~Lange,
\href{https://doi.org/10.1016/S0168-9002(01)00089-4} {Nucl. Instrum. Meth. A {\bf 462}, 152 (2001);}
R.~G.~Ping,
\href{https://doi.org/10.1088/1674-1137/32/8/001}{Chin. Phys. C {\bf 32}, 599 (2008).}


\bibitem{lundcharm}
J. C. Chen, G. S. Huang, X. R. Qi, D. H. Zhang, and Y. S. Zhu,
\href{https://journals.aps.org/prd/abstract/10.1103/PhysRevD.62.034003}{Phys. Rev. D {\bf 62}, 034003 (2000).}

\bibitem{photos}
E.~Richter-Was,
\href{https://doi.org/10.1016/0370-2693(93)90062-M`}{Phys. Lett. B {\bf 303}, 163 (1993).}

\bibitem{Bugg:1994mg}
D.~V.~Bugg, V.~V.~Anisovich, A.~Sarantsev and B.~S.~Zou,
\href{https://journals.aps.org/prd/pdf/10.1103/PhysRevD.50.4412}{Phys. Rev. D {\bf 50}, 4412-4422 (1994).}
\bibitem{Bugg:2008ig}
D.~V.~Bugg,
\href{https://arxiv.org/pdf/0808.2706.pdf}{Phys. Rev. D, {\bf78}, 074023(2008).}

\bibitem{BESIII:2016tqo}
M.~Ablikim \textit{et al.} (BESIII Collaboration),
\href{https://journals.aps.org/prd/abstract/10.1103/PhysRevD.95.032002}{Phys. Rev. D {\bf 95}, 032002 (2017).}

\bibitem{mark3}
R. M. Baltrusaitis {\it et al.} (MARK-III Collaboration),
\href{https://journals.aps.org/prl/pdf/10.1103/PhysRevLett.56.2140}{Phys. Rev. Lett. {\bf 56}, 2140 (1986)};
J. Adler {\it et al.} (MARK-III Collaboration),
\href{https://journals.aps.org/prl/pdf/10.1103/PhysRevLett.60.89}{Phys. Rev. Lett. {\bf 60}, 89 (1988).}


\bibitem{deltakpi}
M. Ablikim {\it et al.} (BESIII Collaboration),
\href{https://doi.org/10.1016/j.physletb.2014.05.071}{Phys. Lett. B {\bf 734}, 227 (2014).}

\bibitem{argus}
H. Albrecht {\it et al.} (ARGUS Collaboration),
\href{https://doi.org/10.1016/0370-2693(90)91293-K}{Phys. Lett. B {\bf 241}, 278 (1990).}

\bibitem{punzi}
G.~Punzi,
\href{https://arxiv.org/pdf/2011.11770.pdf}{arXiv:2011.11770}[physics.data-an].

\bibitem{trokle}
W.~A.~Rolke, A.~M.~Lopez and J.~Conrad,
\href{https://arxiv.org/pdf/physics/0403059.pdf}{Nucl. Instrum. Meth. A \textbf{551}, 493-503 (2005).}

\bibitem{ROOT}
R. Brun and F. Rademakers,
\href{https://doi.org/10.1016/S0168-9002(97)00048-X}{Nucl. Instrum. Meth.  {\bf 389}, 81 (1997).
}

\bibitem{ks}
M. Ablikim {\it et al.} (BESIII Collaboration),
\href{https://doi.org/10.1103/PhysRevD.92.112008} {Phys. Rev. D {\bf 92}, 112008 (2015).}


\bibitem{BESIII:2017pez}
M.~Ablikim \textit{et al.} (BESIII Collaboration),
\href{https://journals.aps.org/prd/abstract/10.1103/PhysRevD.96.111101}{Phys. Rev. D {\bf 96}, 111101 (2017).}

\bibitem{BESIII:2018nzb}
M.~Ablikim \textit{et al.} (BESIII Collaboration),
\href{https://journals.aps.org/prl/abstract/10.1103/PhysRevLett.121.171803}{Phys. Rev. Lett. \textbf{121}, 171803 (2018).}

\bibitem{BESIII:2021uqr}
M.~Ablikim \textit{et al.} (BESIII Collaboration),
\href{https://journals.aps.org/prl/abstract/10.1103/PhysRevLett.127.131801}{Phys. Rev. Lett. {\bf 127},  131801 (2021).}

\bibitem{BESIII:2015hty}
M.~Ablikim \textit{et al.} (BESIII Collaboration),
\href{https://journals.aps.org/prd/abstract/10.1103/PhysRevD.94.032001}{Phys. Rev. D {\bf 94}, 032001 (2016).}

\bibitem{Ke:2023qzc}
B.~C.~Ke, J.~Koponen, H.~B.~Li and Y.~Zheng,
\href{https://www.annualreviews.org/doi/10.1146/annurev-nucl-110222-044046}{Ann. Rev. Nucl. Part. Sci. \textbf{73} 285-314(2023).}

\bibitem{Li:2021iwf}
H.~B.~Li and X.~R.~Lyu,
\href{https://doi.org/10.1093/nsr/nwab181}{Natl. Sci. Rev. \textbf{8} (2021).}



\end{thebibliography}
